\documentstyle[prl,aps]{revtex}
\input epsf
\begin{document}
\draft
\twocolumn[
\widetext
\title{Absence of zero field muon spin relaxation induced by
superconductivity in the B phase of UPt$_3$}
\author{P. Dalmas de R\'eotier, A. Huxley, A. Yaouanc and J. Flouquet}
\address{Commissariat \`a l'Energie Atomique, D\'epartement de Recherche
Fondamentale sur la Mati\`ere Condens\'ee, \\ F-38054 Grenoble Cedex 9,
France}
\author{P. Bonville, P. Imbert and P. Pari}
\address{Commissariat \`a l'Energie Atomique, D\'epartement de Recherche
sur l'Etat Condens\'e, les Atomes et les Mol\'ecules, \\ F-91191
Gif-sur-Yvette, France}
\author{P.C.M. Gubbens and A.M. Mulders}
\address{Interfacultair Reactor Instituut, Delft University of Technology,
2629 JB Delft, The Netherlands}
\date{\today} \maketitle \widetext
\begin{abstract}
\leftskip 54.8pt
\rightskip 54.8pt
We present muon spin relaxation measurements performed on
crystals of the heavy fermion superconductor UPt$_3$. In zero applied
field, contrary to a previous report,  we do not
observe an increase of the internal magnetic field in the lower
superconducting phase (the B phase).
Our result gives an experimental upper bound of the magnetic field that could
be associated with the superconducting state.
\end{abstract}
\pacs{PACS numbers : 74.70.Tx, 75.20.Hr, 76.75.+i}
]
%%%%%%%%%%%%%%%%%%%%%%%%%%%%%%%%%%%%%%%%%%%%%%%%%%%%%%%%%%%%%%%%%%%%%%%%%%%%%%%
\narrowtext
The unconventional nature of superconductivity in the hexagonal heavy fermion
superconductor UPt$_3$ is now well established \cite{Flouquet}.
This was first convincingly demonstrated by the occurence in zero
magnetic field of two superconducting
phase transitions at $T_{\rm {C+}}$ $\simeq$ 0.50 K and $T_{\rm {C*}}$
$\simeq$ 0.45 K.
Despite intense experimental and theoretical activities, the nature of
superconductivity in UPt$_3$ is still unknown. Even the
question of the symmetry of the superconducting order parameter
is not resolved \cite{Sauls,Heffner95}. \par
Recently Luke {\it et al.} have reported zero field muon
spin relaxation ($\mu$SR) measurements which show that the relaxation rate
increases when crossing $T_{\rm {C*}}$ from above \cite{Luke93a,Luke93b}.
This observation has been taken as a proof that the B phase (the low
temperature
superconducting
phase below $T_{\rm {C*}}$) breaks time reversal symmetry and is characterized
by triplet Cooper pairs. This remarkable
result has attracted much interest
\cite{Ohmi,Machida,Sauls,Heid,Heffner95}.\par
The physical properties of heavy fermion metals are well known to be extremely
dependent on the sample quality. High quality crystals are now available.
These two facts have been our original
motivation for performing new $\mu$SR measurements, the results of which are
presented in this letter. \par
The samples were prepared from two large single crystals of UPt$_3$ grown
by the Czochralski method under ultrahigh vacuum from zone-refined
depleted uranium. The as-grown crystals were annealed for several days at
1200 $^\circ$C -1300 $^\circ$C (see Ref.~\onlinecite{Brison}), and
then spark-cut
or wire-sawed. The cut plates were annealed for an additional week
at 950 $^\circ$C, improving the sharpness of the resistivity
transition (an example is presented at Fig. 1 of Ref.~\onlinecite{Keller}).
The resistivity  of our samples, measured on long samples (8 mm length,
0.3 $\times$ 0.4 mm$^2$ cross section) obeys $\rho(T)$ = $\rho(0)$ +
$AT^2$ perfectly below 1.1 K, with $\rho_{\bf c}(0)$ =
0.28 $\mu\Omega$cm, $\rho_{\bf a}(0)$ = 0.51 $\mu\Omega$cm and $A_{\bf c}$ =
0.90 $\mu\Omega$cmK$^{-2}$, $A_{\bf a}$ = 1.46
$\mu\Omega$cmK$^{-2}$ for
current parallel to $\bf c$ and $\bf a$. These low residual
resistivities, which are
comparable with recently published values \cite{Lussier}, are
a proof of the high quality of our samples.\par
The $\mu$SR measurements were carried out on two samples which differ by the
orientation of the crystal axes
relative
to the sample plane : either $\bf c$ or $\bf a^*$ ($\equiv$ $\bf b$)
is perpendicular
to it. Each sample is a disk of $\sim$ 18 mm diameter
and $\sim$ 0.4 mm thickness, comprising of a mosaic of $\sim$ 10
aligned slices of the same single crystal carefully glued to a 5N silver
plate (40 $\times$ 40 mm$^2$).
To ensure good thermal contact at low temperature, each
slice is ultrasonically bonded to a gold wire clamped to the silver plate. This
method allows to reach low temperatures as shown previously
by specific heat and thermal conductivity
measurements \cite{Brison}. \par
The $\mu$SR measurements were performed at the MuSR
spectrometer \cite{Eaton} of the ISIS
surface muon beam facility located at the Rutherford Appleton Laboratory
(RAL, UK). The spectra were recorded with a $^3$He-$^4$He dilution
refrigerator (designed at
the Commissariat \`a l'Energie Atomique, Saclay, France) for
temperatures below 4.2 K and with a helium cryostat (a so called orange
cryostat, Institut Laue Langevin, Grenoble, France)
for temperatures up to 30 K.
Some cross checked spectra were recorded down to 1.75 K with the orange
cryostat.\par
In the $\mu$SR technique polarized muons are implanted into a sample where
their spin evolves in the local magnetic field until they decay
\cite{Chappert,Schenck}. The decay
positron is emitted preferentially along the final muon spin direction; by
collecting several million positrons, we can reconstruct the time dependence of
the muon spin depolarization function $P_Z(t)$ which, in turn, reflects
the distribution of fields experienced at the muon site. The
$Z$ axis refers to the muon beam polarization
axis which, in our case, is as well the direction of the detected positrons
because the measurements have been performed with the longitudinal
geometry \cite{Chappert,Schenck}. $P_Z(t)$ has been deduced from
the raw data using the
method described in Ref. \onlinecite{Dalmas90}.
We have carried out measurements at zero field and with an external applied
field of 10 mT. \par
\begin{figure}
\epsfbox{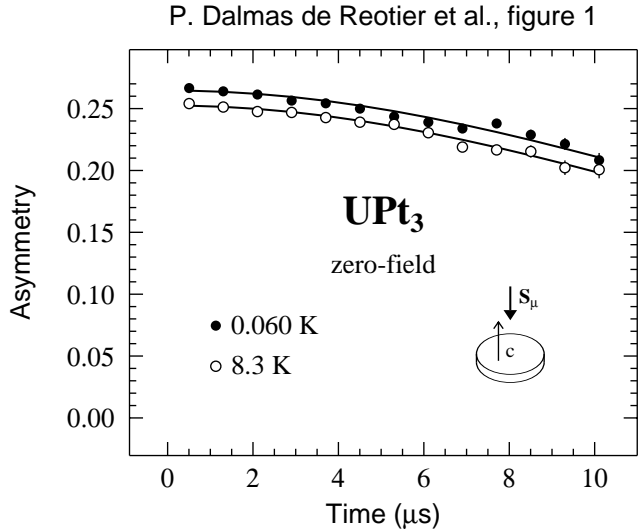}
\caption{Typical zero field spectra measured on the UPt$_3$ sample with the
$c$ axis parallel to the initial muon beam polarization, ${\bf S}_\mu$. The
solid (empty) circles refer to the spectrum recorded below (above) the
superconducting temperatures. The initial asymmetry of the two spectra
is not exactly the same because they have
been recorded with different cryostats. The full lines are fit with
Eq. (\protect \ref{fitsum}). This figure shows that there is no
additional internal
magnetic field in the low temperature superconducting phase. }
\label{spectra_zf}
\end{figure}

In Fig.~\ref{spectra_zf} we present examples of zero field spectra.
They are well analysed by
the sum of two functions
\begin{eqnarray}
P_Z(t)=a_{\text{KT}}P_{\text{KT}}(t) + a_{\text{bg}}\exp(\lambda_{\text{bg}}t),
\label{fitsum}
\end{eqnarray}
where $P_{\text{KT}}(t)$ is the Kubo-Toyabe function which describes the
relaxation due to the sample and the second term accounts for the muons
stopped in the sample holder, cryostat walls and windows. Separate
measurements performed in a
transverse field for the same experimental geometry showed that $a_{\text{KT}}$
$\simeq$ 0.165 and $a_{\text{bg}}$ $\simeq$ 0.100. Measurements at zero
field with
only the silver plate and no sample showed that a good estimate
of $\lambda_{\text{bg}}$
is 0.012 MHz for the present sample size. The UPt$_3$ zero field spectra were
therefore fitted with
$a_{\text{bg}}$ and $\lambda_{\text{bg}}$ fixed to the previous values and
$a_{\text{KT}}$ as a free parameter. $a_{\text{KT}}$ is then found to be
constant over the temperature range investigated.
The Kubo-Toyabe function is written
\begin{eqnarray}
P_{\text{KT}}(t)={1 \over 3} + {2 \over 3} (1-\Delta_{\text{KT}}^2t^2)
\exp(-{1 \over 2} \Delta_{\text{KT}}^2t^2),
\label{KTfunction_full}
\end{eqnarray}
where the Kubo-Toyabe relaxation rate
$\Delta_{\text{KT}} = \gamma_\mu \sqrt {\langle B^2 \rangle}$ describes the
width of the distribution of local fields.
$\gamma_\mu$ is the muon gyromagnetic ratio ($ \gamma_ \mu$ = 851.6
Mrad s$^{ {\rm -1}}$T$^{-1} $) and ${\langle B^2 \rangle}$ the second moment of
the field distribution at the muon site. Because the argument of the
exponential term in Eq. (\ref{KTfunction_full}) is small in our case,
$P_{\text{KT}}(t)$ is well approximated by a parabolic
function :
\begin{eqnarray}
P_{\text{KT}}(t)=1 - \Delta_{\text{KT}}^2t^2.
\label{KTfunction_appro}
\end{eqnarray}
The parabolic character of the spectra is clearly seen in
Fig.~\ref{spectra_zf}. The fact that the depolarization due to
the samples is well described by the
Kubo-Toyabe function is a strong indication that the spins of the muons are
depolarized by a static field distribution \cite{Chappert,Schenck}. This
interpretation is confirmed by additional measurements performed at high
and low temperatures with a magnetic field of 10 mT applied along the $Z$
axis : with this experimental set-up the spectra are not depolarized. \par
\begin{figure}
\epsfbox{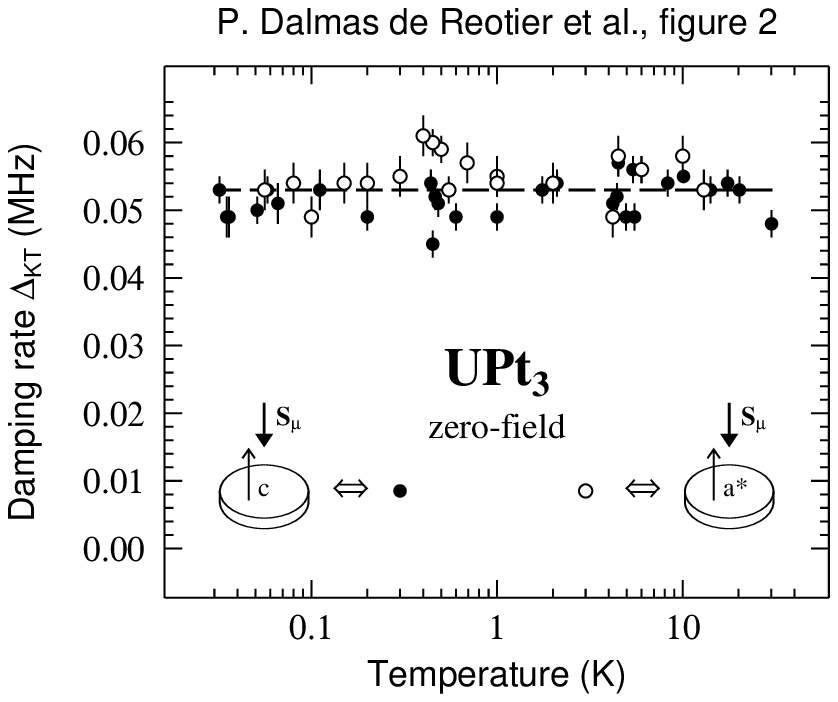}
\caption{Temperature dependence of the Kubo-Toyabe relaxation rate
$\Delta_{\text{KT}}$ measured on two UPt$_3$ samples which differ by the
orientation of ${\bf S}_\mu$ relative to the crystal axes : ${\bf S}_\mu$ is
either parallel to $\bf c$ or $\bf a^*$. The dashed straight
line indicates the average $\Delta_{\text{KT}}$ values. These
results show that $\Delta_{\text{KT}}$
is independent of the temperature and orientation of the crystal axes
relative to ${\bf S}_\mu$.}
\label{delta_zf}
\end{figure}

In Fig.~\ref{delta_zf} we present the temperature dependence of
$\Delta_{\text{KT}}$ for the two orientations of the crystal axes
relative to the $Z$ axis. $\Delta_{\text{KT}}$ is temperature independent for
the two samples. From this observation we deduce that a possible change in the
internal magnetic field by magnetism or superconductivity, if it
exists, has to be smaller that approximately 3 $\mu$T at the muon site over the
whole temperature range.
We note that $\Delta_{\text{KT}}$ has almost the same mean value for the two
samples. The rest of this letter discuss these experimental facts.
We first compare our results with previously published ones.\par
Our measurements
do not detect the antiferromagnetic magnetic phase at $T_N$ $\sim$
5 K first inferred from $\mu$SR measurements \cite{Heffner} and confirmed by
neutron diffraction \cite{Aeppli88}. Neutron diffraction
measurements performed on some of the slices of our $\mu$SR samples show
that the antiferromagnetic phase transition still exists with the same
published characteristics \cite{Fak}. The increase of
$\Delta_{\text{KT}}$ detected by the
authors of Ref. \onlinecite{Heffner} could be a consequence of a modification
of the muon stoping site due to a larger concentration of impurities or defects
in the sample. \par
While Luke {\it et al.} found approximately the same $\Delta_{\text{KT}}$
value as us above $T_{C*}$
\cite{Luke93a,Luke93b}, below that temperature their spectra are more
damped. For example in Ref. \onlinecite{Luke93a} they present six temperature
points characterized by a $\Delta_{\text{KT}}$ increase of $\sim$ 0.01 MHz at
low temperature.
This is definitively not seen in our samples. As above the observed
increase of damping rate was also probably due to the
impurities and/or the defects contained in their samples. Although
Luke {\it et al.} do not
give detailed information on the characteristics of their samples, we have
indirect evidence that they are not of the best quality. We have inferred this
conclusion using the following reasoning : i) additional transverse field
$\mu$SR measurements on our samples show that the transverse
relaxation rates are clearly larger that in the samples of
Luke {\it et
al.} \cite{Futur}; ii) the transverse relaxation rates measured by Luke {\it et
al.} and Broholm {\it et al.} \cite{Broholm} are consistent; iii) the sample of
these latter authors which is described in some detail in Refs.
\onlinecite{Broholm} and \onlinecite{Kleiman} was not annealed, {\it i.e.} it
was not of the best possible quality.\par
Our results show clearly that the muon spins are not depolarized by magnetic
moments of electronic origin. Therefore the observed depolarization is induced
by the nuclear magnetic moments which are almost uniquely carried by the
$^{195}$Pt nuclei of abundance 33.7 \%. Given a muon localization
site, $\Delta_{\text{KT}}^2$ due
to nuclear moments can be computed using the well known
formula \cite{Chappert,Schenck}
\begin{eqnarray}
\Delta_{\text{KT}}^2 & = & {1 \over 6} \sum_i
\left ( {\mu_o \over 4 \pi} {\gamma_\mu \gamma_i \hbar \over r_i^3} \right)^2
\cr
& \times &
I_i \left(I_i + 1 \right) \left( 5 - 3 \cos ^2 \theta_i \right),
\label{Delta}
\end{eqnarray}
where the sum is over the nuclei located at a distance $r_i$ from
the muon. Their gyromagnetic ratio is $\gamma_i$ and spin $I_i$.
$\mu_0$ is the permeability of free space and $\theta_i$ the angle
between the $Z$ axis and ${\bf r}_i$. In our case only 33.7 \% of the Pt nuclei
have a spin
(= 1/2). Therefore $\Delta_{\text{KT}}^2$ = $\left(\sum_N
\Delta_{\text{KT},N}^2 \right)/N$
where the sum is over the $N$ possible $^{195}$Pt nuclei configurations
around the muon. The muon localization site is unknown in UPt$_3$.
But because we do not
observe muon spin oscillations below $T_N$ (not even an increase in damping),
the average magnetic field (dipolar
plus hyperfine) at the muon site must be zero. This result is explained if the
muon sits in
a site of high symmetry such as the octahedral (0,0,1/2) site as found for
the isostructural intermetallic CeAl$_3$ \cite{Barth}. For this site we
find $\Delta_{\text{KT}}$ = 0.061 MHz, 0.061 MHz
and 0.060 MHz for $Z$ parallel to the $\bf a$, $\bf b$ and $\bf c$ axes,
respectively.
Therefore for the considered site,  $\Delta_{\text{KT}}$ is predicted to be
practically
isotropic as found experimentally and has approximately the measured value.
The 10 \% difference between the theoretical and experimental
$\Delta_{\text{KT}}$ values can be attributed to a $\sim$ 3 \% increase of the
distance between the muon and the Pt nearest neighbour atoms as found for
metallic Cu \cite{Camani} and/or to the muon zero point motion
\cite{McMullen}.\par
Interestingly, a precise zero field nuclear magnetic resonance (NMR) experiment
performed on the Pt nuclei at 1.4~K failed to observe any magnetic ordering in
UPt$_3$
\cite{Kohori}. Whereas a symmetry argument may explain the non observation by
$\mu$SR
spectroscopy of the phase transition at $T_N$ (see above), in the NMR case
there is no such
argument. \par
We note that the $\Delta_{\text{KT}}$ value observed at high temperature is
consistent with the one found previously
\cite{Heffner,Luke93a,Luke93b}. The difference only appears at the phase
transitions ($T_N$ and $T_{C*}$). Qualitatively it can be understood if we
suppose that in the previous works the high symmetry ($\bar{3}m$) of
the octahedral interstitial sites was broken by crystal strains and/or defects.
If so, the delicate balance between the magnetic
fields produced by the uranium electronic moments which results in no field at
the muon site is broken.
The observed $\Delta_{\text{KT}}$ increase in Ref.
\onlinecite{Luke93a,Luke93b} could then reflect a possible modification of the
magnetic structure induced by superconductivity in the B phase. Related to
this interpretation we note that
a decrease of $\sim$ 5\% of
the (1,1,1/2) magnetic Bragg peak intensity in the B phase has
already been seen in neutron
scattering \cite{Aeppli89}. \par
The fact that we do not see any
change for $\Delta_{\text{KT}}$ as a function of temperature
supports the idea that the observed magnetic inhomogenity in transverse field
muon experiments carried out for the B phase is only produced by the vortex
lattice as supposed in Ref.~\onlinecite{Kleiman}.\par
We now compare the available theoretical predictions for the magnetic field
induced in the superconducting state of
UPt$_3$ at the muon site, $B_{\text{super}}$,
to the upper bound value found experimentally : $B_{\text{super}}$
$\lesssim$ 3$\mu$T. Undamped
ring currents always produces an orbital field, $B_{\text{orb}}$. Depending on
whether the Cooper pair has a spin or not, an addition field due the spin
density, $B_{\text{spin}}$ has to be added \cite{Mineev}.
$B_{\text{orb}}$ is expected to be smaller than $B_{\text{spin}}$ by a factor
$k_B T_{\rm {C*}}/ \epsilon_F$ where $k_B$ is the Boltzman constant and
$\epsilon_F$ the Fermi energy \cite{Mineev}. For UPt$_3$
Choi and Muzibar predict $B_{\text{orb}}$ = 0.6 $\mu$T \cite{Choi}. As
$k_B T_{\rm {C*}}/ \epsilon_F$ $\simeq$ 2$\times 10^{-3}$ we deduce
$B_{\text{spin}}$ $\simeq$ 300 $\mu$T ($\epsilon_F$ is computed using the
data of Ref. \onlinecite{Chen}). This is consistent with the value
given by
Mineev for UBe$_{13}$ when rescaled for UPt$_3$ \cite{Mineev} and is much
higher than our upper experimental bound.
In summary our results are consistent with singlet Cooper pairs but before
giving a definite conclusion further
theoretical work is required to resolve the ambiguities between the different
internal fields predicted for non singlet states.\par

\acknowledgements{
The researchers from the Netherlands
acknowledge support from the Dutch Scientific Organisation (NWO). We thank J.
Chappert for all the energy he has spent to make the $\mu$SR dilution fridge a
reality at the MuSR spectormeter and the ISIS facility crew for the
excellent working conditions. We are grateful to V.P. Mineev, P. Coleman and A.
Tsvelik for their interest
and comments.
The $\mu$SR measurements were partly supported by the Commission of the
European Community through the Large Installations Plan.}



\begin{references}
\bibitem{Flouquet} J. Flouquet {\it et al.\/}, Physica (Amsterdam) {\bf 185C},
372 (1991); N. Grewe and F. Steglich in {\sl Handbook on the Physics and
Chemistry of Rare Earths}, edited by G. Schneider {\it et al.\/}, Vol. 14
(1991) ; L. Taillefer, Hyperfine Int. {\bf 85}, 379 (1994) and references
therein.
\bibitem{Sauls} J.A. Sauls, Adv. in Phys. {\bf 43}, 113 (1994) and references
therein.
\bibitem{Heffner95} R.H. Heffner and M.R. Norman, Comments Cond. Mat. Phys., to
be published.
\bibitem{Luke93a} G.M. Luke {\it et al.\/}, Physica (Amsterdam)
{\bf 186B}, 264 (1993).
\bibitem{Luke93b} G.M. Luke {\it et al.\/}, Phys. Rev. Lett. {\bf 71},
1466 (1993).
\bibitem{Ohmi} T. Ohmi {\it et al.\/}, Phys. Rev. Lett. {\bf 71},
625 (1993).
\bibitem{Machida} K. Machida {\it et al.\/}, J. Phys. Soc. Jpn {\bf 62}, 3216
(1993).
\bibitem{Heid} R. Heid {\it et al.\/}, Phys. Rev. Lett. {\bf 74},
2571 (1995).
\bibitem{Brison} J.P. Brison {\it et al.\/}, Physica (Amsterdam)
{\bf 199B $ \& $ 200B}, 70
(1994); J.P. Brison {\it et al.\/}, J. Low temp. Phys.
{\bf 95}, 145 (1994).
\bibitem{Keller} N. Keller {\it et al.\/}, Phys. Rev. Lett. {\bf 73},
2364 (1994).
\bibitem{Lussier} B. Lussier {\it et al.\/}, Phys. Rev. Lett. {\bf 73},
3294 (1994).
\bibitem{Eaton} G.H. Eaton {\it et al.\/}, Hyperfine Int. {\bf 85}, 1099
(1994).
\bibitem{Chappert} {\sl Muon and Pions in Material Research}, edited by J.
Chappert and R.I. Grynszpan (North-Holland, Amsterdam, 1984).
\bibitem{Schenck} A. Schenck, {\sl Muon Spin Rotation Spectroscopy\/} (Adam
Hilger, Bristol, 1985).
\bibitem{Dalmas90} P. Dalmas de R\'eotier {\it et al.\/}, Hyperfine Int. {\bf
65}, 1113 (1990).
\bibitem{Heffner} R.H. Heffner {\it et al.\/}, Phys. Rev. B {\bf 39}, 11345
(1989).
\bibitem{Aeppli88} G. Aeppli {\it et al.\/}, Phys. Rev. Lett. {\bf 60},
615 (1988).
\bibitem{Fak} B. F\aa k {\it et al.\/}, unpublished results.
\bibitem{Futur} A. Yaouanc {\it et al.\/}, unpublished results.
\bibitem{Broholm} C. Broholm {\it et al.\/}, Phys. Rev. Lett. {\bf 65},
2062 (1990).
\bibitem{Kleiman} R.N. Kleiman {\it et al.\/}, Phys. Rev. Lett. {\bf 69},
3120 (1992).
\bibitem{Barth} S. Barth {\it et al.\/}, Phys. Rev. B {\bf 39}, 11695 (1989).
\bibitem{Camani} M. Camani {\it et al.\/}, Phys. Rev. Lett. {\bf 39},
836 (1977).
\bibitem{McMullen} T. McMullen and E. Zaremba, Phys. Rev. B {\bf 18}, 3026
(1978).
\bibitem{Kohori} Y. Kohori {\it et al.\/}, J. Mag. Mag. Mat. {\bf 90 $ \& $
 91}, 510 (1990) and K. Asayama private communication.
\bibitem{Aeppli89} G. Aeppli {\it et al.\/}, Phys. Rev. Lett. {\bf 63},
676 (1989).
\bibitem{Mineev} V.P. Mineev, JETP Lett. {\bf 49}, 719 (1989).
\bibitem{Choi} C.H. Choi and P. Muzikar, Phys. Rev. B {\bf 39}, 9664 (1989).
\bibitem{Chen} J.W. Chen {\it et al.\/}, Phys. Rev. B {\bf 30}, 1583 (1984).
%\bibitem{Midgley} P.A. Midgley {\it et al.\/}, Phys. Rev. Lett. {\bf 70},
%678 (1993).
\end{references}
\end{document}